\documentclass{mem}
\usepackage{natbib}\usepackage{txfonts}\usepackage{balance}
\usepackage{graphicx}
\usepackage{txfonts}
\usepackage[a4paper]{hyperref}
\idline{79}{3}
\begin{document}
\def\teff{$T\rm_{eff }$}
\def\kms{$\mathrm {km s}^{-1}$}

\title{
The Tip of the Red Giant Branch
}


\author{
M. \,Bellazzini\inst{1} 
          }

  \offprints{M. Bellazzini}

\institute{
Istituto Nazionale di Astrofisica --
Osservatorio Astronomico di Bologna, Via Ranzani 1,
I-40127 Bologna, Italy
\email{michele.bellazzini@oabo.inaf.it}
}

\authorrunning{Bellazzini}

\titlerunning{The Tip of the Red Giant Branch}

\abstract{
I review the latest results on the calibration of the Tip of the Red Giant
Branch as a standard candle, in the optical and in the near infrared. 
The agreement among different and independent 
empirical calibrations is rather good, if all the uncertainties are
taken into account. The possible extension of the calibration to SDSS
photometric bands ({\tt i,z}) is also discussed. 
}
\maketitle{}

\section{Introduction}

The use of the Tip of the Red Giant Branch (TRGB) as a standard candle is a
mature technique that is currently used to reliably estimate distances to
galaxies of all morphological types, from the
Local Group \cite[see, f.i.,][and references therein]{mcc} up to the Virgo
cluster and beyond \cite[see, f.i.,][and references therein]{ferra,durre}.
The underlying physics is well understood \citep[][hereafter MF98]{scw,mf98}
and the observational procedure is operationally well defined and robust to
contamination by fore/background stars and/or stars brighter than the TRGB
\citep[MF98,][]{sakai}. The key observable is the sharp cut-off occurring at 
the bright end of the Red Giant Branch (RGB) Luminosity Function (LF) that can 
be easily detected with the application of either non-parametric 
\citep{mf95,sakai} or parametric \cite[see, for example][]{mendez,mcc} methods. 
The necessary condition for a safe application of the technique is that the 
observed RGB LF be well populated, with more than $\sim$100 stars 
within 1 mag from the TRGB \citep{mf95,b02}. Under these conditions, the
typical uncertainty on the estimate of the apparent magnitude of the 
tip $I^{TRGB}$  is of order of a few hundreds of mag ($\le 0.05$), and the
uncertainties on $M_I^{TRGB}$ dominates the error budget of distance estimates
obtained with the TRGB technique.
In general, the magnitude of the
TRGB depends quite weakly on the age of the considered population
\citep{bark,scw} and, in the Cousins' I band, it depends weakly also on the
metallicity, at least for relatively metal poor systems \citep{DA90,lfm93}.
In Near Infra Red (NIR) passbands the dependence on metallicity is strong but
it can be accounted for with very simple models 
\cite[see below, and also][and references therein]{f00,iva,ele,b04}.
 
Here, I will briefly review the status of the calibration of this standard
candle and the perspectives for future applications. For obvious reasons of
space this review cannot be and is not intended to be exhaustive or complete; 
for example,
I will not report on the hundreds of applications of the method, 
I will focus on the most
recent results, I will deal more with calibrations in the
optical than in  NIR passbands, and, also, I will not discuss the differences
between the prediction of the various theoretical models \citep{scw}. 
More detailed discussions and
references can be found in MF98, \citet{sc98}, \citet[][hereafter B01]{b01},
\citet[][hereafter B04]{b04}, and \citet[][hereafter R07]{r07}. The TRGB method in the broader
context of  distance scales is discussed by \citet{W03} and \citet{alves},
among the others. The behaviour of the TRGB in complex stellar
populations is investigated by \citet{bark} and  \citet{sg05}. A detailed and
up-to-date report of stellar evolution on the Red Giant Branch is given by
\citet{scw}.

Nature provides serious limitations in our possibility of studying the
behaviour of the TRGB as a function of metallicity, age, elemental abundance
pattern, etc., since we lack populous star clusters covering 
the whole extension of the space of parameters. 
For this reason, to have an
idea of - at least - the differential changes of the TRGB magnitude in a given
passband in response to variations of the above quoted parameters we 
must recur to grids of
theoretical models \cite[which may suffer from several problems and 
unadequacies, see][]{scw}. 
Also in this case a complete overview of all the available
sets of tracks/isochrones would be impossible: 
here I use mainly 
BASTI
models
\citep{basti,cordier}, but also Padova models \citep{g00,g04}, Yale-Yonsei
(Y$^2$) models \citep{YY}, the old-ages isochrones by \citet{scl97}, and, 
finally, the recently published Darthmouth models \citep{dotter}.

\begin{figure}[]
\resizebox{\hsize}{!}{\includegraphics[clip=true]{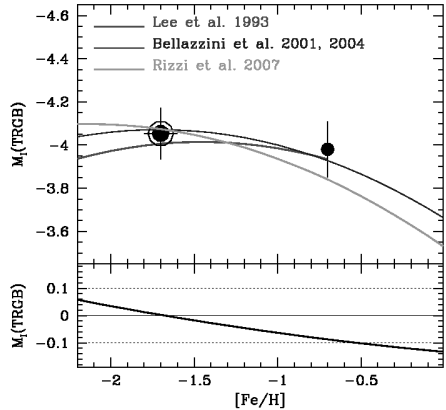}}
\caption{\footnotesize
Comparison between different empirical calibrations of $M_I^{TRGB}$ as a
function of metallicity (upper panel). The large filled circle is 
the observed tip level of $\omega$ Cen, the smaller circle corresponds to the
tip of 47 Tuc, according to B04. The R07 calibration, as a function of $(V-I)_0$
color has been expressed as a function of [Fe/H] using the $(V-I)_0^{TRGB}$ vs.
[Fe/H] relation by B01.
The lower panel shows the difference between the B04 and R07 calibrations as a 
function of [Fe/H].}
\label{compz}
\end{figure}

\section{Empirical calibrations}

Until now, any empirical calibration of the TRGB assumed a model having the
form:

\begin{equation}
M_I^{TRGB}=f([Fe/H])+ZP
\label{mod}
\end{equation}

where $f([Fe/H])$ is a simple function of the metallicity (or of a proxy for
metallicity), typically a polynomial, and ZP is the corresponding Zero 
Point\footnote{Here I make the case for the Cousins' I passband, but the 
same kind of model is used also for calibrations in other bands.}. 
The [Fe/H] range of validity of the calibration must be specified. 
This kind of models implicitly assume that the impact of
other parameters (like age, Helium abundance, abundance pattern etc.) 
on the TRGB magnitude can be
neglected. In fact uncertainties in these parameters can introduce errors in 
the determination of $M_I^{TRGB}$ of order $\pm 0.1$ mag that we are currently
renouncing to account for, also given the difficulty in obtaining observational 
constraints on these parameters, in particular for distant galaxies. For
instance, the uncertainty on the metallicity scale alone (\citet{zw84} versus
\citet{cg97}) may introduce errors in $M_I^{TRGB}$ as large as $\pm 0.05$ mag;
stellar populations having the same total content of heavy elements Z
but differing by $0.3$ dex in 
$[\alpha/Fe]$ may display differences in $M_I^{TRGB}$ as large as $\pm 0.1$ 
mag, depending on the metallicity\footnote{For this reason it is advisable to
use some form of global metallicity, when possible, like for example 
[M/H] \citep[][and references therein]{scs,f99}; see B04.}. 
On the other hand, for ages larger than $\sim
4$ Gyr and $[M/H]\le -0.5$ there is essentially no (reasonable) variation of 
a physical parameter other than metallicity that seems able to produce
variations of $M_I^{TRGB}$ larger than $\pm 0.1$ \cite[][MF98,B04]{bark,sg05}.
In summary, the adoption of the family of models of Eq.~\ref{mod} seems
justified, at the present stage, but the user should be aware of all the 
(possibly) associated uncertainty. Finally, it must be recalled that the Zero
Point itself is known with an uncertainties of $\pm 0.12$ , in the best case
(B04); TRGB estimates of distance moduli with error bars smaller than this
figure neglect part of the actual error budget.
 
\subsection{The calibrating systems}

The natural calibrators of the TRGB method are Galactic Globular Clusters 
(GGC): they
are well studied systems with known age and metallicity \citep{DA90,lfm93}.
The main drawback of globulars is tied to their nature of low luminosity stellar
systems, that means, in most cases, an upper part of the RGB LF that is too
poorly populated to provide reliable detection of the LF cut-off
\citep{cr84,b02}.
In practice there is just a handful of GCs that are (potentially) suitable as
calibrating pillars (B04). B01 tried to circumvent this limitation by taking the
form of $f([Fe/H])$ from a large set of GCCs  from \citet[][F99]{f99} and the 
Zero
Point from the clean detection of the Tip obtained in the most populous GGC,
i.e. $\omega$ Cen (B01). 

\begin{figure}[]
\resizebox{\hsize}{!}{\includegraphics[clip=true]{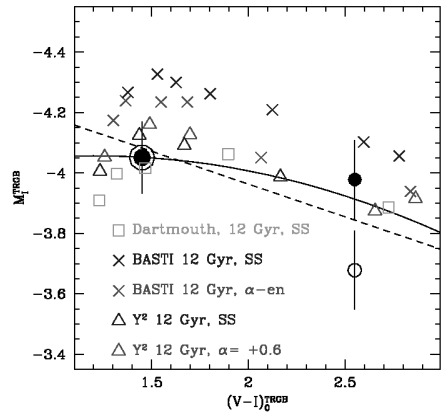}}
\caption{\footnotesize
Comparison of the B01 (continuous line) and R07 (dashed line) calibrations of $M_I^{TRGB}$
as a function of the color of the Tip, and the theoretical predictions of different sets of
theoretical models, with solar scaled (SS) chemical composition and various 
[$\alpha$/Fe] ratios. 
The filled circles are the same as in Fig.~\ref{compz}, above; the open circle is the 
level of the tip of 47 Tuc if the very recent distance estimate by \citet{mac} 
is adopted (see B04 for a discussion on the distance of this cluster).
The difference between the distance of 47~Tuc adopted by B04 and the one
estimated by \citet{mac} gives a quantitative idea of the uncertainties that are
still affecting even the {\em local} distance scale.
}
\label{model}
\end{figure}

A general limitation of any currently available calibration resides in the
uncertainty of the distance to the calibrating systems, usually obtained from RR
Lyrae or (equivalently) from the Horizontal Branch level (F99); this approach
makes the TRGB method a tertiary distance indicator. By adopting the distance
modulus for $\omega$ Cen obtained by \citet{thom} from a double-lined detached
eclipsing binary, B01 tied the ZP of their calibration to a semi-geometrical
distance estimate, thus virtually making the TRGB a primary distance indicator.
The GAIA astrometric satellite \citep{gaia} will measure trigonometric
parallaxes of a large number of $\omega$ Cen stars, thus providing an
exquisitely accurate distance to the cluster. This will provide an iron-clad ZP
for the B01 calibration and will definitely turn the TRGB into a primary 
distance indicator. The B01 calibration has been extended also to NIR
passbands in B04. 

R07 introduced two very interesting novelty in their approach to the empirical
calibration. 
While their distance scale fully relies on the HB, with all the
associated limitations (see B01), they (a) adopt well resolved nearby galaxies as
calibrating systems, and, (b) they calibrate  $M_I^{TRGB}$ as a function of the
color of the tip $(V-I)_0^{TRGB}$, instead of metallicity or of other less well
defined/behaved proxies \citep{lfm93}.
 Point (a) resolve the problem of poorly
populated RGB LF that affects GCs, since the calibrating system are much larger
systems, plenty of RGB stars; moreover they easily cover a large range
in metallicity, in particular reaching solar and super-solar values, out of the
reach of GGCs. The RGB of a large galaxy can be so rich 
of stars that it can be
divided in color strips, obtaining more than one estimate of the TRGB level for
the same galaxy, at different $(V-I)_0^{TRGB}$ (see R07). 
Averaging these estimates would
mitigate possible systematics associated with the actual Star Formation History
(SFH) of the considered system \cite[see][]{sg05}.
Point (b) allows the simultaneous measure of the actual observable
(the  RGB LF cut-off) and a very sensitive metallicity proxy 
(the color at the cut-off) from the same Color Magnitude Diagram (CMD). 
In all cases
in which an estimate of the metallicity of the stellar system under
consideration must be obtained from the CMD 
a TRGB calibration based on $(V-I)_0^{TRGB}$ should be
preferred as it uses a direct observable that does not depend on the distance,  
\cite[see][and references therein, for different approaches to the same 
problem]{lfm93,sakai,b02}.

As the B01/B04 calibration is provided as a function of [Fe/H] and [M/H] while
R07 one is provided as a function of $(V-I)_0^{TRGB}$, some transformation 
is needed to actually compare the two calibrations. 
We use the relations from B01
to transform both calibrating relations. In particular the B01
calibration becomes:

\begin{equation}
M_I^{TRGB}=a(V-I)_0^2-b(V-I)_0-c
\label{vitip}
\end{equation}
where $a=0.080$, $b=-0.194$, and $c=-3.939$, and
$(V-I)_0$ is the color at the tip, i.e. $(V-I)_0^{TRGB}$.

In Fig.~\ref{compz} different empirical calibrations of $M_I^{TRGB}$
(expressed as a function of [Fe/H]) are compared. 
The overall agreement is reasonably good. In
particular B01/B04 and R07 calibration are completely independent and are based
on completely different sets of calibrating systems. 
Still, the two calibrations
are in excellent agreement (within $\simeq \pm 0.05$ mag for $[Fe/H]\le -1.0$;
the difference becomes larger than $\simeq 0.1$ mag only for 
$[Fe/H]\ge -0.5$). In the $M_I^{TRGB}$ vs. $(V-I)_0^{TRGB}$ plane
(Fig.~\ref{model}) the agreement is even better. 
The comparison with various sets
of models shown in Fig.~\ref{model} suggests two main conclusions: 
(1) the observed slope is reasonably
reproduced by all the models, over a very large color range; (2) there is a large
variation in the predicted ZP, that may be partly due to the color
transformations adopted in the various models (see B04 for a discussion).
A deeper discussion of differences among the various theoretical models is
clearly beyond the scope of this paper \citep[see][]{scw}. Limited preliminary
experiments suggests that different models may provide significantly different
predictions for the variations of $M_I^{TRGB}$ in response to variations of
$[\alpha/Fe]$; the issue is not treated by existing theoretical studies and it
probably deserves a deeper investigation. 

\subsection{Helium abundance}

In recent years it has emerged the possibility that the early chemical evolution
of (at least some) stellar system may lead to the generation of stars with a
very high abundance of Helium, up to $Y\sim 0.40$ 
\citep[see][and these proceedings]{norr,dan,pio,sol}, triggering a renewed
interest in the study of the effects of He abundance on remarkable features of 
CMDs.
\begin{figure}[]
\resizebox{\hsize}{!}{\includegraphics[clip=true]{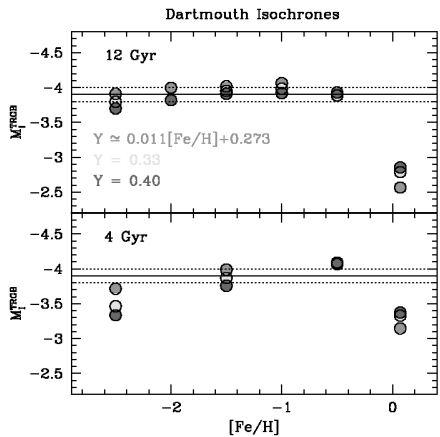}}
\caption{\footnotesize
$M_I^{TRGB}$ as a function of [Fe/H] for models with different Helium abundances
Y. Upper panel: 12 Gyr old models; Lower panel: 4 Gyr old models. 
}
\label{hel}
\end{figure}
\begin{figure}[]
\resizebox{\hsize}{!}{\includegraphics[clip=true]{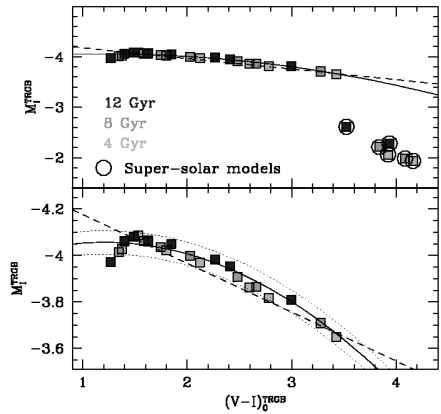}}
\caption{\footnotesize
Upper panel: $M_I^{TRGB}$ as a function of the color of the tip
$(V-I)_0^{TRGB}$ for solar-scaled  BASTI models of different ages (4 - 8 - 12
Gyr, corresponding to light - medium - dark grey filled squares; all models
have been shifted by +0.24 mag in $M_I$ to match the empirical calibrations,
see Fig.~\ref{model}, B01 and B04).  The large empty circles marks models with
solar and  super-solar metallicity. The continuous line is the B01 calibration
in the form of Eq.~\ref{vitip}, the dashed line is the R07 calibration. Lower
panel: the same as above but zoomed on the models with sub-solar metallicity.
The symbols are the same as in the upper panel. 
The dotted lines enclose the $\pm 0.05$
mag range around the B01 calibration.
}
\label{etavi}
\end{figure}

In Fig.~\ref{hel} I use Darthmout models to explore the behaviour of 
$M_I^{TRGB}$ in response to {\em large} variations of Y. For old populations, 
the
effect is relatively small. It is interesting to note the inversion of the sense
of the dependence at the extremes of the metallicity range: at $[Fe/H]=-2.5$ the
population with the lowest Y has the brightest Tip, the opposite is true at 
solar metallicity. 

Fig.~\ref{hel} suggests that at intermediate ages ($\sim$ 4 Gyr) the effect of 
Helium on $M_I^{TRGB}$ may be very strong. The issue certainly deserve to be
investigated in deeper detail with other sets of theoretical models. In any
case, it should be recalled that Fig.~\ref{hel} explores quite extreme regimes
of He abundances (far larger than the Sun, for instance), and a conclusive case
for the existence of such extremely-He-rich stars (at least in large numbers) 
is yet to be done.

\subsection{Age, high metallicities and NIR.}

In Fig.~\ref{etavi} BASTI models of different ages are superposed to the
B01/B04 and R07 calibrations, after a +0.24 mag shift to match the ZP of the
empirical calibrations. The lower panel shows that, 
for $(V-I)_0^{TRGB}\le 3.5$, the effect of age variations
on $M_I^{TRGB}$ is within $\le \pm 0.05$ mag, thus
confirming the weakness of the age dependency, in the range 4 Gyr - 12 Gyr . 
The upper panel shows that for
solar and super-solar populations the dependency of $M_I^{TRGB}$ on color (and
metallicity) becomes strong and may also require more complex models.

\begin{figure*}[t!]
\resizebox{\hsize}{!}{\includegraphics[clip=true]{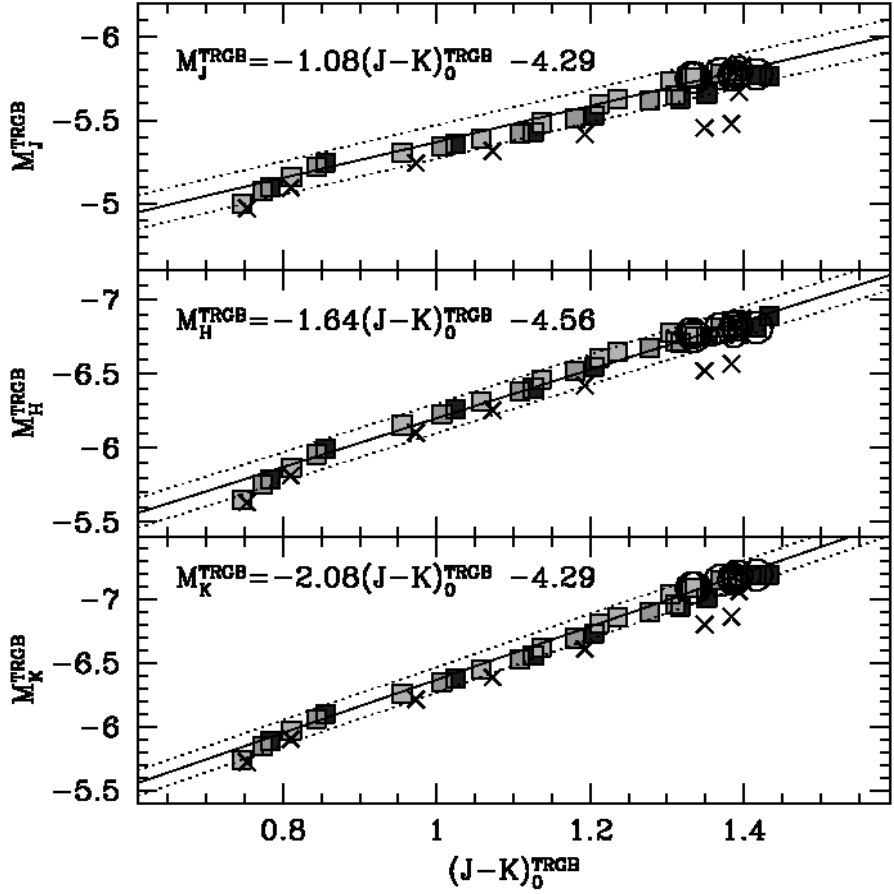}}
\caption{\footnotesize
J,H,K absolute TRGB magnitudes as a function of $(J-K)_0^{TRGB}$ color from
the  BASTI models for different ages; the symbols are the same as in
Fig.~\ref{etavi}, with the addition of the $\times$ symbols that are
$\alpha$-enhanced, 12 Gyr old models.
The continuous lines are derived from the empirical
calibrations by B04; the  dotted lines enclose the $\pm 0.1$ mag range around
the B04 calibration.}
\label{nir}
\end{figure*}

This problem is largely removed by passing to NIR photometry. In Fig.~\ref{nir}
the empirical calibrations of the absolute J,H,K magnitude of the tip as a 
function of  $(J-K)_0^{TRGB}$ obtained from B04 data are compared to BASTI
models of different ages and [$\alpha$/Fe] ratios. While the dependence on color
(metallicity) is quite strong, a simple linear model provide a good description
of the TRGB magnitude to $\simeq \pm 0.1$ mag, up to super-solar metallicities.
As the most powerful and innovative telescopes of the future 
(JWST\footnote{\tt www.jwst.nasa.gov/}, 
ELT\footnote{\tt www.eso.org/projects/e-elt/}, 
TMT\footnote{\tt www.tmt.org/}, etc.)
will be dedicated to / optimized for infrared observations and they will
probably allow us to resolve the RGB of very metal rich elliptical
galaxies, the importance of  NIR calibrations of the TRGB will be ever growing
in the next couple of decades.

In Fig.~5 the calibrating equations for the absolute magnitude of the Tip as a
function of $(J-K)_0^{TRGB}$ color instead of [M/H] are also reported, as
derived from B04 data.

\subsection{The TRGB in the SDSS system}

\begin{figure}[]
\resizebox{\hsize}{!}{\includegraphics[clip=true]{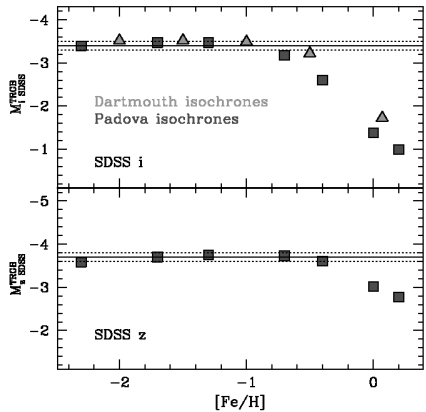}}
\caption{\footnotesize
{\tt i} and {\tt z} absolute TRGB magnitudes as a function of metallicity for 
Dartmouth (triangles) and
Padova (squares) models. The horizontal lines marks $M_i^{TRGB}=-3.44\pm 0.1$ (continuous
and dotted lines, respectively), and $M_z^{TRGB}=-3.67\pm 0.1$.}
\label{sdss}
\end{figure}

The Sloan Digital Sky Survey is providing accurate photometry of a very wide 
portion
of the Northern sky in its own
specific ({\tt ugriz}) photometric system \cite[][]{fuku,adel}. Other currently
ongoing (or planned) large surveys will further extend the sky coverage 
in the same (or very similar) photometric system 
(SEGUE\footnote{\tt segue.uchicago.edu/}, 
PanSTARRS\footnote{\tt pan-starrs.ifa.hawaii.edu/public/}, 
LSST\footnote{\tt www.lsst.org}). It is very likely that most photometric
studies in the future will be performed in this system, as a large number of
(relatively faint) secondary standard stars will be present in any given field
of the sky.
In Fig.~\ref{sdss} we use two sets of models to show that in 
both the {\tt i} and {\tt z} SDSS passbands, the magnitude of the TRGB has a very
weak dependence on metallicity over wide metallicity ranges (as in Cousins' I,
or even better). 
As a preliminary calibration,  $M_i^{TRGB}=-3.44\pm 0.1$ for [Fe/H]$\le -1.0$,
and $M_z^{TRGB}=-3.67\pm 0.1$ for [Fe/H]$\le -0.4$ can be adopted, from
Fig.~\ref{sdss}; it is quite clear that an empirical calibration is needed,
also for these passbands. It is worth noting that PanSTARRS will observe also in
a band (Y) that is intermediate between {\tt z} and J, that may even turn out to
be the most indicated for the calibration of the TRGB as a standard candle (see
R07).

\section{Conclusions}

The future of the TRGB as a distance indicator looks very promising, as new
telescopes and surveys will give access to more distant systems, larger
areas of the sky and more favourable wavelenght windows. In general applications
to distant galaxies, the lack of knowledge of many parameters that may
influence the TRGB magnitudes will probably prevent the method to reach
exquisite levels of accuracy (i.e., better than 10\% uncertainties); 
on the other hand the {\em weakness} of all these dependencies ensures a good
level of robustness and reliability of the method. The Zero Point of current
empirical calibrations is uncertain at the $\pm 0.1$ level, but the results 
the GAIA astrometric mission should reduce such uncertainty 
virtually to zero. An extension of the empirical calibration to passbands like
SDSS {\tt i} and {\tt z} and a detailed comparison between the prediction of the
different sets of theoretical models are suggested as the most interesting
problems to afford, at the present stage.  
 
\begin{acknowledgements}
M.B acknowledges the financial support to this research by INAF, through the
grant CRA 1.06.08.02., assigned to the project {\em A hierarchical merging tale 
told by stars: motions, ages and chemical compositions within structures and
substructures of the Milky Way}.
\end{acknowledgements}

\bibliographystyle{aa}

\end{document}